\documentclass[epj]{svjour}
\usepackage{graphics}

\usepackage{amssymb}
\usepackage{graphicx}
\usepackage{dcolumn}
\usepackage{bm}

\providecommand{\Journal}[4] {#1 {\textbf {#2}}, #3 (#4)}
\providecommand{\PRD}{Phys. Rev. D } %
\begin{document}
\title{The CDF dijet excess from intrinsic quarks}

\title{The CDF dijet excess from intrinsic quarks}

\author{Xiao-Gang He\inst{1,2,3}
\thanks{\emph{email:} hexg@phys.ntu.edu.tw} \and Bo-Qiang Ma\inst{1}% etc
\thanks{Corresponding author. \emph{email:} mabq@pku.edu.cn}%
}                     % Do not remove

\institute{School of Physics and State Key Laboratory of Nuclear
Physics and Technology, Peking University, Beijing 100871 \and
Department of Physics and Center for Theoretical Sciences, National
Taiwan University, Taipei 10617 \and Institute of Particle Physics
and Cosmology, Department of Physics, Shanghai JiaoTong University,
Shanghai 200240}

% \and Center for High Energy Physics, Peking University, Beijing 100871, China}
%

\date{Received: date / Revised version: date}

\abstract{ The CDF collaboration reported an excess in the
production of two jets in association with a $W$. We discuss
constraints on possible new particle state interpretations of this
excess. The fact of no statistically significant deviation from the
SM expectation for {$Z$+dijet} events in CDF data disfavors the new
particle explanation. We show that the nucleon intrinsic strange
quarks provide an important contribution to the $W$ boson production
in association with a single top quark production. Such {$W$+t}
single top quark production can contribute to the CDF {$W$+dijet}
excess, thus the nucleon intrinsic quarks can provide a possible
explanation to the CDF excess in {$W$+dijet} but not in {$Z$+dijet}
events.
\PACS{
      {12.39.-x}{Phenomenological quark models}   \and
      {13.85.Hd}{Inelastic scattering: many-particle final states}  \and
      {14.20.Dh}{Protons and neutrons} \and
      {14.65.Ha}{Top quarks}
     } % end of PACS codes
} %end of abstract
\maketitle
%
%%%%%%%%%%%%%%%%%%%%%%%%%

%\section{Introduction}
%\\

Recently, the CDF collaboration reported an excess in the production
of two jets in association with a $W$ boson production~\cite{CDF},
in the Fermilab Tevatron collider of proton and antiproton collision
at a center-of-mass energy of 1.96~{TeV/c$^2$} with an integrated
luminosity of 4.3~fb$^{-1}$. The $W$ boson is identified through a
charged lepton (electron or muon) with large transverse momentum,
and the invariant mass of the two-jet system ($M_{jj}$) is found to
be in the range of 120-160~{GeV/c$^2$}. The two-jet system appears
to be an unidentified resonance with mass around 150~{GeV/c$^2$},
and this triggers the speculations to understand the dijet as a
non-standard new particle resonance with the associated $W$
production cross section to be about 4~pb. It is also
reported~\cite{CDF} that such an excess of dijet events cannot be
described within the statistical and systematic uncertainties of
current theoretical predictions. Therefore the CDF excess has the
potential of being an indication for new physics beyond the standard
model~(SM).

The purpose of this letter is to make an analysis of some possible
explanations for the CDF dijet excess within and beyond the SM. We
analyse constraints on the properties of the speculated Higgs or
$Z'$ resonance, if one tries to explain the CDF dijet excess as a
signal from a new kind of particles beyond SM. The fact that no
statistically significant deviation from the SM expectation for $Z$
plus dijets events disfavors new particle resonance
explanation~\cite{cdf-bb}. This leads us go back to examine
uncertainties concerning strange and heavy quark content of the
nucleon. We show that some intriguing features of the nucleon sea,
which are relevant to understand several anomalies previously found
in experiments, can also provide some understanding for the newly
reported CDF dijet excess.

%\section{New particle state interpretation}

There are many possible ways a new particle can manifest itself in
theoretical models. Here we briefly discuss some properties of a
beyond SM Higgs boson and a new vector gauge boson with a mass of
150~GeV/c$^2$ for the explanation of the CDF {$W$+dijet} excess.

The CDF {$W$+dijet} excess has been shown to be not compatible with
SM $WH$ production, {\it i.e.}, interpreting the dijet being from
the decay of a SM Higgs with a mass of 150~{GeV/c$^2$}. The main
reason is that the production cross section $\sigma$ multiplied with
the branching ratio ${\mathrm{BR}}(H \to b\bar b)$ of a Higgs decays
into a pair of $b\bar b$ is only about $12$~fb which would not give
excess for $WH \to l\nu b\bar b$. This has been confirmed by the
dedicated CDF search~\cite{cdf-bb}.

When going beyond the SM, it is possible to have a larger branching
ratio for $H \to b \bar b$ and also a larger $WH$ production cross
section. For example, in MSSM there are two Higgs doublets $H_u$ and
$H_d$ giving up- and down-quark masses, respectively. The Yukawa
coupling of $b$ quark to the SM-like Higgs is scaled by a factor
$\tan\beta = v_u/v_d$. Here $v_i$ is the vacuum expectation values
of $H_i$. In the MSSM $\tan\beta$ can be as large as a few tens. The
$\sigma\cdot {\mathrm{BR}}(H\to b\bar b)$ can be enhanced. The same
can be said for many multi-Higgs models. The enhanced coupling also
helps to make the total cross section enhanced. These models also
have charged Higgs contributions. If the mass of the charged Higgs
is not too far away from the neutral ones, they can also enhance the
total dijet excess. However, a universal enhancement in down quark
sector may not be enough to have a large enough $WH$ production
cross section. To this end we note that there are models where the
light quark Yukawa couplings can be enhanced even more, such as the
private Higgs model and other Higgs multiplet
models~\cite{he,he-b,he-c}. In this type of models the total
{$W$+dijet} excess may become more close to the CDF value. This type
of models usually also have large $ZH$ production which may be a
problem. We will come back to this later.

The {$W$+dijet} excess may also be due to new gauge bosons, such as
a $Z'$. There are stringent constraints on the mass of a new $Z'$
gauge boson if it has significant coupling to charged leptons.
Various data, for example the Tevatron dilepton data, have
constrained the mass to be much larger than the 150~{GeV/c$^2$}
resonant mass. However, if the $Z'$ couples primarily to
quarks~\cite{leptoforbic,leptoforbic-b,leptoforbic-c}, that is the
$Z'$ is hadrophilic or leptophobic, the constraints are much less
severe. It is possible to have a low $Z'$ mass to produce the
{$W$+dijet} excess. As we are studying this possibility, several
papers~\cite{kingman,kingman-b,kingman-c,kingman-d} appeared showing
that this is indeed possible if the coupling of the $Z'$ to quarks
is similar in order compared with the $U(1)_Y$ coupling. With such a
large coupling, s-channel production of $Z'$ can also become
significant, but the signal may be buried under the large QCD
background. With a hadrophilic $W'$, it is also possible to explain
the dijet excess.

In the above two classes of models, and also most of models relying
on a new particle resonance~\cite{kingman,tech} to explain the CDF
dijet excess data also predict excess for associated
{($\gamma$,$Z$)+dijet} production. The reason is that the vertex at
which a $W$ is emitted from a quark line, a replacement of $W$ by a
$Z$ or a $\gamma$ is possible. If one starts with the same quark
inside a proton, after the replacement, the quark which emits the
new particle changes identity compared with the $W$ emission. This
may change the cross section significantly if the new particle
couples to up- and down-types of quarks differently. However, there
is also contribution from a different type of quark inside a proton
after the replacement of $W$ by a $Z$ or a $\gamma$ the quark
emitting the new particle is the same type of quark. Since both up-
and down-types of quarks have sizeable non vanishing parton
distribution functions~(PDF) inside a proton, the associated
productions with a $W$ and a $Z$ or a $\gamma$ should have similar
orders of magnitude. If there are other types of vertices, such as
two gauge bosons to a new Higgs or a new gauge boson, the
replacement of a $W$ by a $Z$ in the final state is also expected to
have similar magnitude in strength. This is a generic prediction for
a new particle interpretation of the CDF {$W$+dijet} excess.

In fact CDF has also investigated the shape of the dijet mass
distribution in {$Z$+jets} event. There is no statistically
significant deviation from the SM expectation~\cite{CDF}. This
disfavors the new particle interpretation. At present, one cannot
rule out/in models with high confidence using this data. But this is
something one needs to keep in mind when making claims. Future
improved data can provide with more information. In the following we
discuss an alternative possibility which is free from this potential
problem. This has to do with dijet background from single top quark
production.

%\section{Intriguing features of the nucleon}

There have been several interesting discoveries concerning some
intriguing features of the nucleon which are beyond naive
theoretical predictions. The difference between the EMC measured
value of the spin-dependent structure function of the
proton~\cite{EMC82} and previous theoretical predictions triggered
the proton spin ``crisis" or spin ``puzzle". It has been known that
the relativistic effect due to transversal motions of
quarks~\cite{ma,ma-b,ma-c,ma-d,ma-e} and the sea content of the
nucleon~\cite{Ma:2001ui} play important roles to understand the
proton spin problem. A number of models concerning the sea content
of the nucleon, such as the baryon-meson fluctuation
model~\cite{bm96} and the chiral quark model~\cite{mg84,mg84-b}, not
only play significant roles to understand the proton spin
structure~\cite{Ma:2001ui,bm96,cl95}, but also show their remarkable
significance to explain the flavor asymmetry of the nucleon
sea~\cite{bm96,ehq92,Ding:2006ud,Song:2010rq} reported by the NMC
collaboration~\cite{nmc91,nmc91-b}.
%Such models can produce an excess of
%$d\bar{d}$ pairs over $u\bar{u}$ pairs inside the proton sea, to
%explain the Gottfried sum rule violation found by the NMC
%collaboration~\cite{nmc91}.
More interestingly, the baryon-meson fluctuation model and the
chiral quark model have been also found to produce a
strange-antistrange asymmetry of the nucleon, and such strangeness
asymmetry with natural model estimations~\cite{dm04,dxm04,dxm04b}
can explain the NuTeV anomaly~\cite{zell02,zell02-b} of the
deviation of the NuTeV measured value of weak mixing angle compared
with other measurements.
%This anomaly was attributed as a possible evidence for new physics beyond
%standard model. However, it can be also naturally explained by the
%strange-antistrange asymmetry of the nucleon within the baryon-meson
%fluctuation model~\cite{dm04} and the chiral quark
%$model~\cite{dxm04,dxm04b}.
From the above discussion, we see that the intriguing features of
the nucleon sea have played significant roles in the understanding
of several anomalies beyond theoretical predictions.

Therefore it is natural to examine the possible relevance of the
nucleon sea with the CDF excess in {$W$+dijet} production. The
energy carried away by the produced $W$ boson should be larger than
100~{GeV/c$^2$}, and the dijet should carry more than 200~GeV/c$^2$
of energy. Thus the total energy of the produced {$W$+dijet} event
should be larger than 300~GeV/c$^2$, which means that more than 1/6
of the total energy of the colliding proton and antiproton should be
spent to produce such an event. This happens when at least one of
the parton in the colliding proton and antiproton carried a momentum
fraction larger than 0.1, {\it i.e.}, with Bjorken variable $x \ge
0.1$, a region where the intrinsic quarks (aniquarks) of the nucleon
dominant.

It is important to distinguish between two distinct types of quark
and gluon contributions to the nucleon sea measured in various deep
inelastic processes: ``extrinsic" and ``intrinsic". The sea quarks
generated from the QCD hard bremsstrahlung and gluon-splitting are
referred as ``extrinsic" quarks, since the sea quark structure is
associated with the internal composition of gluons, rather than the
nucleon itself. In contrast, sea quarks which are multi-connected to
the valence quarks of the nucleon are referred to as ¡°intrinsic¡±
sea quarks. It has been shown~\cite{bm96} that the intrinsic
quark-antiquark pairs generated by the minimal energy
nonperturbative meson-baryon fluctuations in the nucleon sea provide
a consistent framework to understand the nucleon intrinsic quarks.
The model predicts an excess of intrinsic $d\bar{d}$ pairs over
$u\bar{u}$ pairs, as supported by the Gottfried sum rule
violation~\cite{Ding:2006ud}. Furthermore, the meson-baryon
fluctuations of the nucleon sea produce a striking quark/antiquark
asymmetry in the momentum distributions for the nucleon strangeness,
and such asymmetry provides a natural explanation~\cite{dm04} of the
NuTeV anomaly within the standard model. The intrinsic quarks of the
nucleon sea can be also alteratively modeled by the chiral quark
model~\cite{Song:2010rq,dxm04}, and all of the above mentioned
anomalies can be also understood as well.

In Fock state wavefunctions containing heavy quarks, the minimal
energy configuration occurs when the constituents have similar
rapidities. Thus one of the most natural features of intrinsic heavy
sea quarks is their contribution to the nucleon structure functions
at large $x$ in contrast to the small $x$ heavy quark distributions
predicted from photon-gluon fusion processes. This feature of
intrinsic charm~\cite{Bro81,Bro81-b,Vog96,Pumplin:2007wg} has been
extensively discussed. The extension of the intrinsic quark idea to
the intrinsic bottom has been also studied~\cite{ib}. The intrinsic
bottom can contribute to the production of a $W$ boson plus a single
top which can decay into two jets~\cite{Heinson}. Therefore the
intrinsic quarks of the nucleon do have relevance to the CDF excess
of {$W$+dijet} events.

%\section{The intrinsic quark contribution to {W+t} production}

The situation can be illustrated in Fig.1, where an intrinsic bottom
(anti-bottom) quark in the colliding proton (or antiproton)
transforms into a $W$ boson and a top (anti-top) quark when
interacting with a gluon in another colliding particle. The produced
top can contribute to the dijet detected by CDF. As some particles
might be missing in the final re-constructed jets, the CDF excess of
dijets in the range 120-160~{GeV/c$^2$} does not exclude the
contribution from decaying of a single top into two jets. The
predicted rate for {$W$+$t$} single top quark production at the
Tevatron collision energy of 1.96~{TeV/c$^2$} is $0.28\pm 0.06$~pb
for a top quark mass of 172.5~{GeV/c$^2$}~\cite{Heinson}. Such a
contribution has been considered as a single top background in the
CDF analysis of {$W$+dijet} events.

\begin{figure}%[htbp]
\centering
\includegraphics[width=0.35\textwidth]{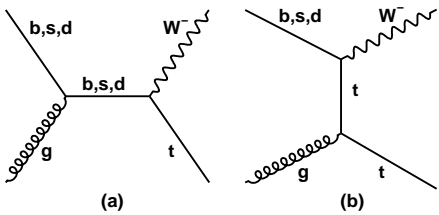}
\caption{\label{fig:ratio} The diagrams for the sub-processes of
single top quark production associated with a $W$ boson production.}
%:(a) the intrinsic bottom quark $b$ of a colliding nucleon interacts
%with a gluon $g$ from another colliding nucleon and then the
%scattered bottom quark $b$ transforms into a $W$ boson and a single
%top quark $t$; (b) the intrinsic bottom quark $b$ from a colliding
%nucleon transforms into a $W$ boson and a single top quark which
%interacts with a gluon from another colliding nucleon. The $b$ quark
%in the diagrams can be replaced by a $d$ or an $s$ quark as well.}
\end{figure}

The intrinsic light-flavor down-type quarks, {\it i.e.}, $d$ and $s$
quarks of the nucleon intrinsic sea might also contribute to the
{$W$+$t$} single top quark production if one replaces the $b$ quark
in Fig.1 by a $d$ or an $s$ quark. Such contributions might be
highly suppressed due to the rather small values of quark mixing
elements $V_{td}=0.00874^{+0.00026}_{-0.00037}$ and
$V_{ts}=0.0407\pm0.0010$ compared with
$V_{tb}=0.999133^{+0.000044}_{-0.000043}$~\cite{pdg}. However, as
the probability of finding intrinsic bottom quarks is also highly
suppressed compared with those of finding $d$ and $s$ quarks, we
need to check their relative ratios between each other from rough
estimates. Unlike ``extrinsic" quarks, which are rather ``soft" at
small $x$ due to productions of the QCD hard bremsstrahlung and
gluon-splitting, the ``intrinsic" quarks of the nucleon can be
rather ``hard" at larger $x$. As it is rather difficult to
distinguish between $d$ and $s$ quarks in the extraction of
flavor-dependent quark distributions, and also the intrinsic $s$
quark distribution is predicted to be harder than the intrinsic
$\bar{s}$ quark distribution in the baryon-meson fluctuation model,
the available parametrizations of parton distribution functions may
have underestimated the strange quark distribution at larger $x$.
Therefore we need to estimate the relative contributions of $d$, $s$
and $t$ quarks from qualitative analysis.

The relative ratios of finding intrinsic $s$ and $b$ quarks inside a
proton can be estimated from their off-shell-ness from the ground
nucleon~\cite{Bro81,Bro81-b}, therefore
\begin{equation}
\frac{\mathrm{Probablity}(s)}{\mathrm{Probablity}(b)} \sim
\frac{m_b^2}{m_s^2}=\left(\frac{4.67~\mathrm{GeV}}{101~\mathrm{MeV}}\right)^2
\approx 2\times {10}^3.
\end{equation}
The relative suppression of $s$ versus $b$ contributions to
{$W$+$t$} production due to the quark mixing can be also estimated
\begin{equation}
\left(\frac{V_{ts}}{V_{tb}}\right)^2=\left(\frac{0.0407}{0.999133}\right)^2\approx
1.7\times {10}^{-3}.
\end{equation}
Therefore the ratios of $s$ versus $b$ quark contribution to
{$W$+$t$} production is 3.6, which corresponds a rate for {$W$+$t$}
single top quark production from $s$ quark contribution of being
1~pb. As $s(x)/\bar{s}(x)$ ranges between $1.5\to 5$ for $x \ge 0.1$
(See fig.5 of Ref.~\cite{dxm04b}), we can therefore suspect that the
$s$ quark contribution is dominant than $\bar{s}$ quark for the
{$W$+$t$} single top production for the colliding proton. This may
also enhance the intrinsic strange quark contribution to the
{$W$+$t$} production by a factor of $2\to 3$, corresponding to $2\to
3$~pb. Considering that the above estimate is very rough, should be
reasonable in the order of magnitude, we thus can conclude that the
intrinsic strange quark of the nucleon has a non-negligible
contribution to the CDF excess of {$W$+dijet} events.
%Our above analysis can only be served as a first qualitative estimate, and
%more sophisticated calculation is still needed for more reliable
%conclusions.

We need also an estimate on the contribution from $d$ quarks. There
is an additional valence $d$ quark inside the proton besides the
intrinsic $s$ quarks at larger $x$. As the intrinsic $s$ quark is of
1/3 of the sea $d$ quark distribution as in previous data
analysis~\cite{dxm04b}, we can thus use a larger probability of 5
times of finding a $d$ quark than an $s$ quark at larger $x$. The
$d$ quark contribution to {$W$+$t$} production is highly suppressed
due to the quark mixing
\begin{equation}
\left(\frac{V_{td}}{V_{tb}}\right)^2=\left(\frac{0.00874}{0.999133}\right)^2\approx
7.7\times {10}^{-5}.
\end{equation}
The rate for {$W$+$t$} single top quark production from $d$ quark
contribution is thus estimated to be 0.16~pb, which is one half
compared with the intrinsic bottom contribution. Thus the $d$ quark
contribution to {$W$+$t$} single top quark production should be also
included as a background in the Tevatron {$W$+dijet} analysis.

%\section{Prediction and discussion}

%From our analysis in the above section, we find that the intrinsic
%strange contribution to {W+t} single top quark production is more
%important than the intrinsic bottom contribution. Our knowledge of
%intrinsic strange quarks is still not fully determined, thus
%uncertainties due to the nucleon intrinsic sea still need further
%investigation from both theoretical and experimental studies.

For the intrinsic strange quarks, the strange-antistrange asymmetry
is a remarkable feature which is predicted from theory~\cite{bm96}
and is further found to provide a viable
explanation~\cite{dm04,dxm04,dxm04b} of the NuTeV anomaly. Here we
find that the strange quark has an important contribution to the
{$W$+$t$} single top production at the Tevatron collider of proton
and antiproton collision. Thus the Tevatron {$W$+$t$} production can
also offer an ideal process to study the strange-antistrange
asymmetry of the nucleon sea with unique advantages. As strange
quarks are more ``hard" than antistrange quarks inside the proton,
we thus can predict that the {$W^-$+$t$} events with $W^{-}$ boson
plus single top quark production should occur more likely in the
proton forward direction, whereas the {$W^+$+$\bar{t}$} events with
$W^{+}$ boson plus single antitop quark production should occur more
likely in the antiproton forward direction. We thus can use the
{($W^-$+t)}/{($W^+$+$\bar{t}$)} asymmetry to extract information on
the strange-antistrange asymmetry of the nucleon. Of course, a  more
quantitative analysis is needed for confrontation with experimental
observation.

One interesting feature of the CDF results is that there is no
anomaly in the dijet mass distribution of {$Z$+jets} events. A
reasonable explanation for the CDF excess of {$W$+dijet} events
should be able to accommodate this feature also. Our suggestion to
use the intrinsic strange contribution to {$W$+$t$} single top quark
production for the {$W$+dijet} excess can accommodate this feature
naturally. There is no quark-flavor-changing process in $Z$
production, therefore no single top production can be resulted from
intrinsic strange and bottom quarks. Thus no excess of dijets with
mass around 150~GeV/c$^2$ can be found in associated {$Z$+jets}
production in our explanation. This is a distinctive feature than
those new particle resonance models. More precise data in future can
distinguish different mechanisms for the CDF dijet excess.

%\section{Summary}
In summary, we analysed the CDF observed excess in the production of
two jets in association with a $W$. If this excess is interpreted as
associated $W$ production with a new particle of mass 150~GeV/c$^2$,
such as a Higgs or a $Z'$ gauge boson, which decays into two jets,
there should also exist dijets excess with an associated $Z$
production. CDF data, however, do not show statistically significant
deviation from the SM expectation for $Z$ plus dijets events. This
disfavors the new particle explanation. This therefore led us to
examine the nucleon intrinsic quark contributions to the {$W$+$t$}
single top quark production. We found that the intrinsic strange
quarks can provide an important contribution to the {$W$+$t$} single
top quark production, with no enhancement in the production of
{$Z$+jets} events. The {$W$+$t$} production contributes to the CDF
{$W$+dijet} events, thus we provide a possible explanation for the
CDF excess of {$W$+dijet} events. We also provide a prediction of
{($W^-$+$t$)}/{($W^+$+$\bar{t}$)} asymmetry in the Tevatron proton
and antiproton collision. Thus our suggestion of a larger intrinsic
strange contribution to the {$W$+$t$} single top quark production
can be tested by detailed analysis of data.

%\begin{acknowledgments}
%\section*{Acknowledgments}
\noindent {\bf Acknowledgments} This work is partially supported by
NSC, NCTS, NSFC (Grants No.~11021092, No.~10975003, No.~11035003,
and No.~11120101004), SJTU 985 grant. BQM acknowledges the support
of NCTS North and warm hospitality from W.-Y. Pauchy Hwang and
Pei-Ming Ho during his visit of NTU.
%, where this work is done.
We also acknowledge the stimulating discussions with Sang Pyo Kim
and Miao Li.

%\end{acknowledgments}

%\section*{Note Added}
%After finishing this work, the CDF reported an updated
%analysis~\cite{CDF2} using data collected through to November 2010
%corresponding to an integrated luminosity of 7.3 fb$^{-1}$. Their
%results are consistent with their early analysis~\cite{CDF} and
%increased the significance to 4.1$\sigma$. Recently D0 collaboration
%also reported their results of an analysis~\cite{Abazov} with an
%integrated luminosity of 4.3 fb$^{-1}$. They did not find similar
%$W+$dijet excess. Although D0 was also looking at similar excess,
%the methodology differs in some way which may be potentially
%important cause for differences. We are not in a position to decide
%which one may be correct which has to be settled among the
%experimental groups. We think that a study of implications of the
%CDF results is still worthy.

\end{document}